\input{aipcheck}

\documentclass[
    ,final            
  ]
  {aipproc}

\layoutstyle{6x9}

\begin{document}

\title{GRB 021004: A Possible Shell Nebula around 
a Wolf-Rayet Star Gamma-Ray Burst Progenitor}

\author{N.\ Mirabal}{
  address={Astronomy Department, Columbia University, 550 West 120th Street,
 New York, NY 10027}
}

\author{
J. P.\ Halpern}{
  address={Astronomy Department, Columbia University, 550 West 120th Street,
 New York, NY 10027}
}

\author{R. Chornock}{
  address={
Department of Astronomy, 601 Campbell Hall, University
        of California, Berkeley, CA 94720-3411}}

\author{A. V. Filippenko}{
  address={
Department of Astronomy, 601 Campbell Hall, University
        of California, Berkeley, CA 94720-3411}}

\iftrue
\author{D. M. Terndrup}{
  address={
Department of Astronomy, Ohio State University, Columbus, OH 43210}}

\begin{abstract}
The rapid localization of GRB 021004 by the HETE-2 
satellite allowed nearly continuous monitoring of its early 
optical afterglow decay, as well 
as high-quality optical spectra that determined a redshift of $z=2.328$ 
for its host, an active starburst galaxy with strong 
Lyman-$\alpha$ emission and several absorption lines. 
Spectral observations show multiple absorbers blueshifted by up to 
3,155 km~s$^{-1}$ relative to the host galaxy Lyman-$\alpha$ emission.
We argue that these correspond to a fragmented shell nebula, gradually 
enriched by a Wolf-Rayet wind over the lifetime of a massive
progenitor bubble. In this scenario, the absorbers can be explained by 
circumstellar material that have been 
radiatively accelerated 
by the GRB emission.
Dynamical and photoionization models are used to provide constraints on the 
radiative acceleration from the early afterglow.
\end{abstract}

\date{\today}

\maketitle

\section{Introduction}

Gamma-ray bursts (GRBs) have been a challenge for 
astronomers ever since their 
serendipitous discovery by the Vela satellites in the late 1960s 
\cite{Kle:1973}.
However, evidence collected over the past six years 
now links ``long-duration'' ($>$ 2 s) GRBs  to the deaths of massive 
stars. Some of the clearest information about the nature of GRBs
comes from the coincidence
of the unusual GRB 980425 with SN 1998bw \cite{Galama:1998},  
and the discovery of 
the Type-Ic supernova, SN 2003dh, nearly simultaneous with GRB 030329
\cite{Stanek:2003}. The temporal coincidence of these events proves 
that long-duration GRBs are associated with 
peculiar Type-Ic supernovae (SNe), and thus are a 
consequence of the evolution of massive stars \cite{Hjorth:2003}. 
The evidence also 
strongly supports the collapsar model for GRBs where a rotating 
massive star (typically a Wolf-Rayet star) undergoes core collapse 
to a black hole surrounded by an accretion disk wind 
\cite{Woosley:1993}. 

While it is now generally accepted that some long-duration GRBs 
are associated with the deaths of massive stars, considerable
uncertainty remains as to what the precise nature of the 
progenitor star is. Theories of stellar evolution suggest
that massive stars lose a large fraction of their mass through 
strong stellar winds \cite{Cassinelli:1979}. 
As a result, one expects to find a substantial amount of
circumstellar material in the 
vicinity of long-duration GRBs. 
Indeed, observations of some GRB environments show compatibility with  
a wind-like medium \cite{Chevalier:1999}.  
There have also been reports of  
strong UV absorption lines in GRB afterglows with outflow velocities
of up to 4,260 km~s$^{-1}$ 
\cite{Barth:2003,Mirabal:2003,Fox:2003}. We discuss how these 
observations can provide information on the GRB progenitor.  

\begin{figure}
  \includegraphics[height=.4\textheight]{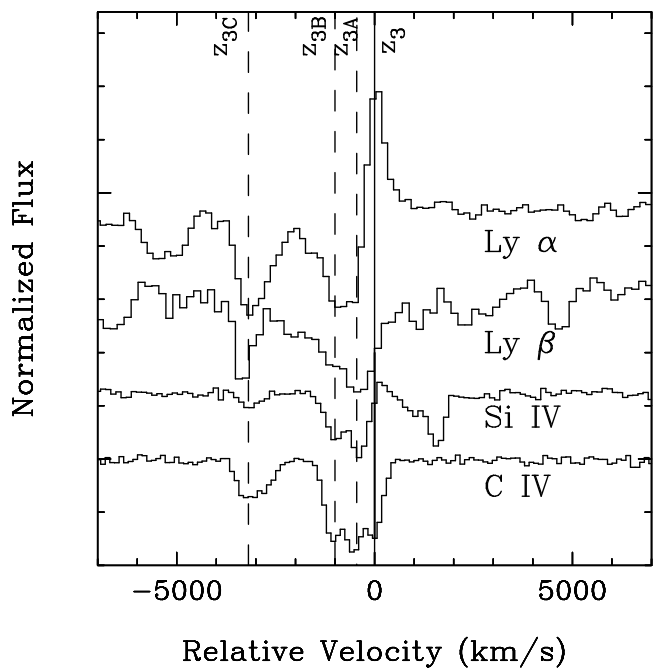}
  \caption{
Outflowing Ly $\alpha$, Ly $\beta$, 
C IV, and Si IV  absorbers in the GRB 021004 
afterglow spectrum plotted 
in velocity space.  As zero velocity we use the systemic redshift 
of the host galaxy $z=2.328$. 
The dashed lines indicate blueshifted absorbers
 at 450, 990, and 3,155 km~s$^{-1}$.}
\end{figure}

\section{Spectral Observations of the GRB 021004 Afterglow}

Possibly the best example of velocity shifts in GRB afterglows is
the optical spectrum of GRB 021004, an active starburst galaxy with strong 
Ly-$\alpha$ emission and several absorption lines.
Spectral observations of its afterglow revealed multiple
blueshifted kinematic components at
 $z_{3A}= 2.323$, $z_{3B}= 2.317$, and $z_{3C}= 2.293$ with
radial velocities of $\sim$ 450, $\sim$ 990, and  $\sim$ 3,155 km~s$^{-1}$ 
relative to the systemic velocity of the host galaxy (Figure 1). 
The absorption components also show velocity widths 
broader than the expected thermal widths at the instrumental resolution,
indicating internal motions within each component. 
Such  
velocity structure is highly unusual for large-scale absorbing material  
near or around the GRB host galaxy. 

One is thus led to consider scenarios 
where the absorbers are closer to the GRB progenitor system (i.e., 
associated). 
Hot, massive stars generally have expanding material 
characterized by velocities of up to 3,000 km~s$^{-1}$, 
which originates through the scattering of stellar radiation in the
stellar wind. Radial outflows 
seen via blueshifted resonance lines are also 
a common trait of a large fraction of Seyfert galaxies 
\cite{Crenshaw:1999}. These
are thought to represent massive outflows of highly ionized gas
from their active galactic nuclei.  
Although there are marked differences between Seyferts and GRBs, 
the similarities in their spectra may shed light on the 
physical conditions of the blueshifted components. 
Building off this premise, we explore a scenario  
where the absorbers in the GRB 021004 spectrum are the result of 
outflowing stellar material associated with the GRB progenitor. 

\section{Shells, Filaments and Clumps}

\begin{figure}
 \includegraphics[height=.3\textheight]{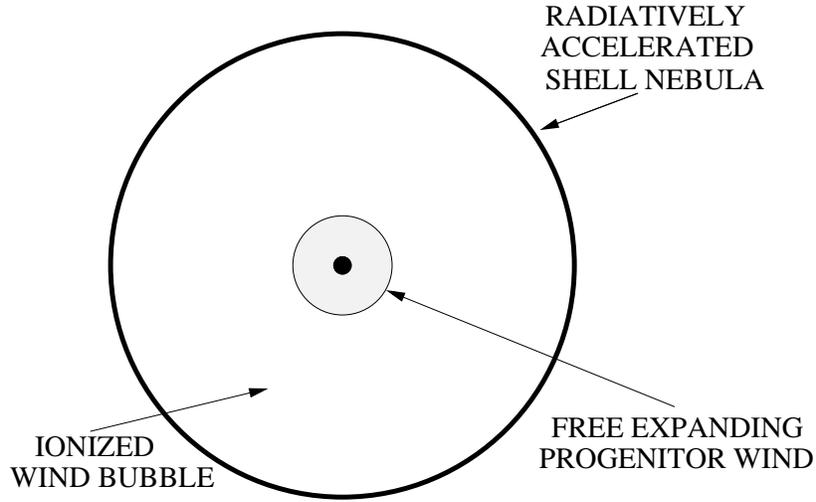}
  \caption{
Schematic cross-section of a stellar-wind bubble model with 
various features including the termination of the wind
and shell nebula. The model cannot reproduce the great wealth of 
structure observed around massive stars.}
\end{figure}

One of the most interesting consequences of 
stellar evolution is that stellar-wind bubbles 
are intimately connected to the 
mass-loss history of their central star. 
On its way to the Wolf-Rayet phase, a main-sequence
star is thought to evolve through a luminous blue variable (LBV) or
red supergiant (RSG) stage.
The slow winds (10--50 km~s$^{-1}$) generated during the LBV or RSG phase
expand into the interior of the main-sequence bubble
until the mass driven by the wind is comparable to
the mass of circumstellar material. 
This condition sets the characteristic
radius of expansion $R$ roughly given by
\begin{equation}
R = \sqrt{\dot{M} \tau \over n_{0}}~{\rm pc}.
\end{equation}
where $\tau$ is wind lifetime in units of 10$^{6}$ years,
 $\dot{M}$ is the mass-loss rate in units of 
10$^{-6}{\rm M}_{\odot}$~yr$^{-1}$, and $n_{0}$ corresponds to the 
density of the surroundings in units of cm$^{-3}$.

Interestingly, stellar winds carry not only mass but kinetic
energy into the ambient medium. 
Such injection of energy leads naturally to 
the formation of overdense shell nebulae 
(with expansion velocities v $\approx$ 
40 km~s$^{-1}$) along the wind profile. Figure 2 shows the predicted
physical structure of a shell nebula 
formed at the termination of a massive stellar wind.
Soon after entering its Wolf-Rayet phase, a
fast wind (1000--3000 km~s$^{-1}$) 
starts sweeping the LBV or RSG material, eventually
overtaking the main-sequence gas residing around 
the progenitor star. The combination of 
streaming winds and internal instabilities within the wind
results in a complex morphology characterized by fragmented shells, 
filaments and clumps. Additional structure is likely to be
introduced by wind-wind collisions in close-binary systems.
But the basic picture is confirmed   
through the variety of morphologies observed in the surroundings of 
isolated Wolf-Rayet stars \cite{Marston:1997}.

Apart from providing a complex circumstellar 
environment, a wind bubble configuration either isolated or in
a close-binary system will 
contain a significant amount of mass from all past stellar phases. 
If GRBs are indeed formed by the death of massive stars, 
overdense stellar regions within the bubble
may produce detectable spectral features in the 
optical afterglow. From the absence
of N V and the presence of Ly $\alpha$, Si IV and C IV,
we inferred that the absorbing
material in the GRB 021004 spectrum was dominated by He-burning and 
core nucleosynthesis 
products. Such composition is in good agreement with  
a massive late-type carbon-rich Wolf-Rayet 
star embedded inside an interstellar
bubble, in which the Wolf-Rayet wind has gradually enriched the 
bubble interior. So far the argument is consistent with the
abundances, but let us explore the kinematics. 

\section{Photoionization and Dynamical Models}

As part of the analysis, 
we modeled the radiation environment in a
wind bubble system following a GRB explosion 
using detailed photoionization models. The models are
especially
constructed to provide constraints on the physical conditions
of blueshifted absorbers. For these particular 
simulations, we  
used the photoionization code IONIZEIT described in \cite{Mirabal:2002},
which includes time-dependent 
photoionization processes taking place under 
a predetermined GRB afterglow ionizing flux.
Perhaps the most dramatic outcome of the simulations is that
material intrinsic to the GRB progenitor should be 
subjected to increasing velocity and gradual ionization. 

In order to explain the absence of significant variability following the first
afterglow spectrum, we concluded that 
radiation acceleration by bound-free transitions had to be 
most efficient in the early stages of the 
GRB \cite{Mirabal:2003, Schaefer:2003}. Figure 3 shows 
the predicted velocity profile for one set of initial
conditions. Faster outflowing absorbers 
can be explained by nearby circumstellar material, while
more distant absorbers will evolve slower.
After final velocity is achieved the absorbers simply coast along. 
In particular, the simulations are able to reproduce 
the kinematics of GRB 021004 if  
high-density clouds are placed at a distance  0.3 $< d <$ 30 pc
from the GRB. The constraints from these models are consistent with the
typical sizes of stellar winds around Wolf-Rayet stars in our Galaxy, and 
provide quite possibly the first direct spectral signature of 
circumstellar material around a GRB progenitor. They also 
lend support to the collapsar model for GRBs where a rotating 
massive star undergoes core collapse 
to a black hole \cite{Woosley:1993}.

\begin{figure}
  \includegraphics[height=.4\textheight]{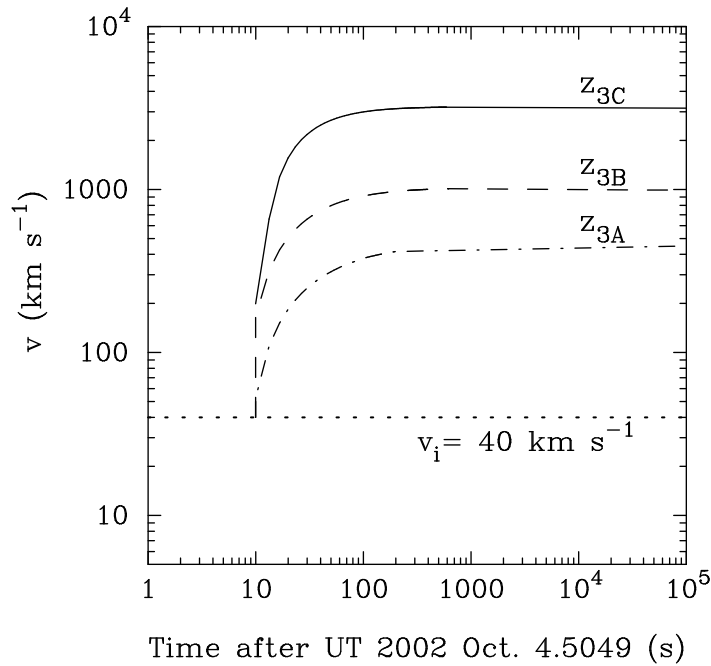}
  \caption{
Simulated velocity profiles for radiatively-accelerated 
outflows.
The dotted line corresponds to an initial outflowing 
velocity of  v $\approx 40$ km~s$^{-1}$.
}
\end{figure}


\section{Conclusions and Future Work}

Observations of the early afterglow of GRB021004 show multiple absorption 
features blueshifted by up to 3,155 km~s$^{-1}$
relative to the host galaxy of the GRB. The features may indeed have 
been caused by 
circumstellar material from a Wolf-Rayet progenitor wind 
located at a distance  0.3 $< d <$ 30 pc from the GRB site 
that has been radiatively accelerated by the GRB afterglow 
emission. While at this stage 
we cannot distinguish between an isolated Wolf-Rayet star and 
a close-binary system, the observational data on GRB 021004 
could be the first direct
spectral signature of material in the surroundings of a GRB.
Our findings motivate the need to undertake
intensive surveys for variable and accelerating resonance lines
(C IV, Si IV,
N V, and O VI), along with Ly $\alpha$ and lower ionization species. 
In addition to the observations, intensive 
numerical modeling is required to deal with the overionization
that takes place when the progenitor gas is exposed to strong GRB emission.
Lastly, we note that the advent of the {\it Swift}
GRB mission (see http://swift.gsfc.nasa.gov/)  
should bring unique access to
early multiwavelength observations of GRBs, which will
provide a critical diagnostic tool for GRB progenitors. 


\IfFileExists{\jobname.bbl}{}
 {\typeout{}
  \typeout{******************************************}
  \typeout{** Please run "bibtex \jobname" to optain}
  \typeout{** the bibliography and then re-run LaTeX}
  \typeout{** twice to fix the references!}
  \typeout{******************************************}
  \typeout{}
 }

\end{document}

\endinput